\documentclass[aps,prl,twocolumn,showpacs,amsmath,superscriptaddress,psfig,amssymb]{revtex4}
\usepackage{amsmath}
\usepackage{amssymb}
\usepackage{graphicx}
\usepackage{dcolumn}
\usepackage{bm}
\usepackage{txfonts}
\usepackage{cases}
\usepackage{color}
\begin{document}
\title{Two-gap superconductivity in LaNiGa$_2$ with non-unitary triplet pairing and even parity gap symmetry}
\author{Z. F. Weng}
\affiliation{Center for Correlated Matter and Department of Physics, Zhejiang University, Hangzhou 310058, China}
\author{J. L. Zhang}
\affiliation{Center for Correlated Matter and Department of Physics, Zhejiang University, Hangzhou 310058, China}
\author{M. Smidman}
\affiliation{Center for Correlated Matter and Department of Physics, Zhejiang University, Hangzhou 310058, China}
\author{T. Shang}
\affiliation{Center for Correlated Matter and Department of Physics, Zhejiang University, Hangzhou 310058, China}
\author{J. Quintanilla}
\affiliation{SEPnet and Hubbard Theory Consortium, University of Kent, Canterbury CT2 7NH, United Kingdom}
\author{J. F. Annett}
\affiliation{H. H. Wills Physics Laboratory, University of Bristol, Tyndall Avenue, Bristol BS8 1TL, United Kingdom}
\author{M. Nicklas}
\affiliation{Max Planck Institute for Chemical Physics of Solids,D-01187 Dresden, Germany}
\author{G. M. Pang}
\affiliation{Center for Correlated Matter and Department of Physics, Zhejiang University, Hangzhou 310058, China}
\author{L. Jiao}
\affiliation{Center for Correlated Matter and Department of Physics, Zhejiang University, Hangzhou 310058, China}
\author{W. B. Jiang}
\affiliation{Center for Correlated Matter and Department of Physics, Zhejiang University, Hangzhou 310058, China}
\author{Y. Chen}
\affiliation{Center for Correlated Matter and Department of Physics, Zhejiang University, Hangzhou 310058, China}
\author{F. Steglich}
\affiliation{Center for Correlated Matter and Department of Physics, Zhejiang University, Hangzhou 310058, China}
\affiliation{Max Planck Institute for Chemical Physics of Solids,D-01187 Dresden, Germany}
\author{H. Q. Yuan}
\email{hqyuan@zju.edu.cn}
\affiliation{Center for Correlated Matter and Department of Physics, Zhejiang University, Hangzhou 310058, China}
\affiliation{Collaborative Innovation Center of Advanced Microstructures, Nanjing 210093, China}
\date{\today}

\begin{abstract}
The nature of the pairing states of superconducting  LaNiC$_2$ and  LaNiGa$_2$ has to date remained a puzzling question. Broken time reversal symmetry has been observed in both compounds and a group theoretical analysis implies a non-unitary triplet pairing state. However all the allowed non-unitary triplet states have nodal gap functions but most thermodynamic and NMR measurements indicate fully gapped superconductivity in LaNiC$_2$. Here we probe the gap symmetry of LaNiGa$_2$ by measuring the London penetration depth, specific heat and upper critical field. These measurements demonstrate two-gap nodeless superconductivity in LaNiGa$_2$, suggesting that this is a common feature of both compounds. These results allow us to propose a novel triplet superconducting state, where the pairing occurs between electrons of the same spin, but on different orbitals. In this case the superconducting wavefunction has a triplet spin component but isotropic even parity gap symmetry, yet the overall wavefunction remains antisymmetric under particle exchange. This model leads to a nodeless two-gap superconducting state which breaks time reversal symmetry, and therefore accounts well for the seemingly contradictory experimental results.

\end{abstract}

\pacs{74.70.Dd; 74.25.Bt; 74.20.Rp}
\maketitle

The breaking of  symmetries in addition to gauge symmetry upon entering the superconducting state usually indicates an unconventional order parameter. Several materials have been found to break time reversal symmetry (TRS) in the superconducting state through the detection of spontaneous magnetic fields below $T_c$ using zero-field muon-spin relaxation ($\mu$SR). In some cases, such as Sr$_2$RuO$_4$ \citep{80_time-reversal_1998} and UPt$_3$ \citep{PhysRevLett.71.1466, UPt3NoTRS} where TRS breaking is also supported by measurements of the polar Kerr effect \cite{PhysRevLett.97.167002,Schemm11072014}, there exists additional  evidence for  triplet superconductivity \citep{Sr2RuO4NMR,133RevModPhys.75.657,RevModPhys.74.235,PhysRevLett.80.3129}. Recently, other superconductors have been reported to show TRS breaking, such as Re$_6$Zr \citep{PhysRevLett.112.107002} and Lu$_5$Rh$_6$Sn$_{18}$ \citep{Lu5Rh6Sn18}, but there is not yet other evidence for unconventional superconductivity and fully gapped behavior is observed. In general the breaking of TRS does not necessarily imply triplet pairing and it is expected for some multiband singlet states such as $s+is$, where there is a  phase difference between the gaps which is neither zero or $\pi$ \citep{PhysRevB.60.14868}. However a particular conundrum is presented by the TRS breaking in LaNiC$_2$ \citep{135PhysRevLett.102.117007} and LaNiGa$_2$ \citep{137PhysRevLett.109.097001}, where it has been argued that as a result of the low symmetry of the orthorhombic crystal structures of both compounds, broken TRS necessarily implies non-unitary triplet superconductivity and all the TRS breaking states have nodes in the gap function \citep{PhysRevB.82.174511}. Although  evidence for nodal superconductivity was found from some measurements  \citep{Lee1996138,NewJPhys13123022}, recent  specific heat \citep{PhysRevB.58.497,NewJ.Phys.15.053005}, nuclear quadrapole relaxation \citep{Iwamoto1998439} and penetration depth \citep{NewJ.Phys.15.053005} measurements indicate fully gapped behavior in LaNiC$_2$. In addition,  evidence for two-gap superconductivity was found from the specific heat, superfluid density and upper critical field \citep{NewJ.Phys.15.053005}. There have been fewer measurements of superconductivity in LaNiGa$_2$ \citep{LaNiGaRep}, which has an orthorhombic centrosymmetric crystal structure in contrast to noncentrosymmetric LaNiC$_2$, although fully gapped behavior was inferred from the specific heat \citep{136PhysRevB.66.092503}.

In this Letter, we suggest a solution to this apparent contradiction from measurements of the London penetration depth, specific heat and upper critical field, all of which consistently suggest the presence of two-gap superconductivity in LaNiGa$_2$. Along with previous results of LaNiC$_2$ \citep{NewJ.Phys.15.053005}, we establish that nodeless, two-gap superconductivity is a common feature of these compounds. We propose that pairing between electrons with the same spins but on different orbitals  gives rise to a triplet superconducting state with even parity pairing in both compounds, where the wave function remains antisymmetric overall due to a sign change upon exchanging electrons between different orbitals. Here additional lowering of the free energy is achieved by an additional field that splits the spin-up and spin-down Fermi surfaces, leading to two distinct gap values.

Polycrystalline LaNiGa$_2$ samples were prepared by arc melting stoichiometric quantities  of La (99.98\%), Ni (99.99\%) and Ga (99.999\%)  in argon gas. The ingots were sealed in evacuated quartz tubes and annealed at 600$^{\circ}$C for one month. Powder x-ray diffraction measurements showed that the samples are single phase with lattice parameters consistent with previous results \cite{136PhysRevB.66.092503}. The residual resistivity of $\rho_0~\approx~1.6~\mu\Omega~$cm and $RRR = \rho_{300K} / \rho_{4K} \approx 28$ indicate a high sample quality and a transition temperature $T_c \approx 1.8$~K was determined from the onset of a sharp superconducting transition. The ac magnetic susceptibility was measured in a $^3$He cryostat and heat capacity measurements were performed using Quantum Design Physical Property Measurement System (PPMS). The London penetration depth was measured in a $^3$He cryostat ($0.4~$K$<T<3~$K) and a dilution refrigerator ($0.05$~K$<T<0.8~$K) utilizing a tunnel diode oscillator (TDO) based technique, where the change of the penetration depth is proportional to the TDO frequency shift, i.e., $\Delta\lambda(T)=\lambda(T)-\lambda_0=G\Delta f(T)$, where $\lambda_0$ is the zero temperature penetration depth and $G$ is solely determined by the coil and sample geometry \cite{GfactorPRB}.

\begin{figure}[t]
\centering
\includegraphics[width=7.0cm]{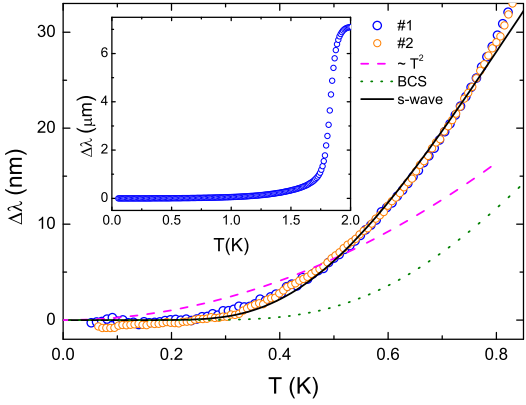}
\caption{(Color online) Temperature dependence of the penetration depth $\Delta\lambda(T)$ of two samples of LaNiGa$_2$ at low temperatures. The solid and dashed lines show fits of $\Delta\lambda(T)$ to an $s$-wave model with  $\Delta(0)=1.30k_BT_c$, and a $T^2$ dependence respectively. The dotted line shows the behavior of an isotropic, weakly coupled BCS superconductor. The inset shows $\Delta\lambda(T)$ for sample $\#$1 up to 2~K.} \label{fig1}
\end{figure}

Figure~\ref{fig1} shows the temperature dependence of the change in the London penetration depth [$\Delta\lambda(T)$] for two samples of LaNiGa$_2$, where  $G$ is $5.2~{\AA}$/Hz and $11.6~{\AA}$/Hz for sample $\#$1 and $\#$2, respectively. The inset displays $\Delta\lambda(T)$ from 2~K down to 0.05~K for sample $\#$1. The signal drops abruptly around the transition temperature $T_c=1.8$~K, which is consistent with the $T_c$ from resistivity (not shown) and ac susceptibility measurements (inset of Fig.~\ref{fig4}). The low temperature data of $\Delta\lambda(T)$ is displayed in the main panel of Fig.~\ref{fig1}. For nodal superconductors at low temperatures, $\Delta\lambda(T)$  shows power law behavior $\Delta\lambda(T)\sim T^n$, with $n=1$ for line nodes and $n=2$ for point nodes. Our data does not display either of these behaviors and the flattening of $\Delta\lambda(T)$ indicates nodeless superconductivity in LaNiGa$_2$. For isotropic $s$-wave superconductors at $T<<T_c$, $\Delta\lambda=\lambda_0\sqrt{\pi\Delta(0)/2k_BT}$exp$[-\Delta(0)/k_BT]$, where $\Delta(0)$ is the zero temperature gap amplitude. As shown by the solid line in Fig.~\ref{fig1}, the data is well fitted by this expression at low-temperatures. The fitted gap of $\Delta(0)=1.30k_BT_c$ is significantly smaller than the weakly-coupled BCS  value of $1.76k_BT_c$, indicating either multiple gaps or gap anisotropy. The behavior of $\Delta\lambda(T)$ for such a BCS model (dotted line) shows poor agreement.

\begin{figure}[t]
\centering
\includegraphics[width=7.0cm]{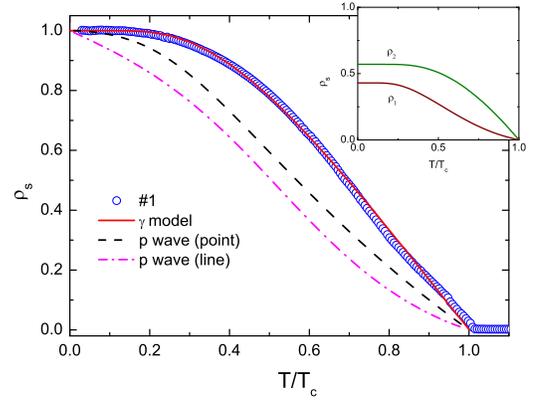}
\caption{(Color online) Superfluid density $\rho_s(T)$ against $T/T_c$. The solid line shows the fitted two-band model, while the dashed and dashed-dotted lines show models with point and  line nodes respectively. The inset shows the components of the two-band model.} \label{fig2}
\end{figure}

To further analyze the gap symmetry of LaNiGa$_2$, we calculated the superfluid density using $\rho_s(T)=[\lambda_0/\lambda(T)]^2$; where $\lambda_0=350$~nm is from $\mu$SR experiments \cite{137PhysRevLett.109.097001}. Figure~\ref{fig2} shows the superfluid density $\rho_s(T)$ for sample \#1 where the flat behavior at low temperatures again indicates fully gapped superconductivity. The superfluid density is shown for two nodal gap structures which cannot account for the data, a model with point nodes where $\Delta_k=\Delta(T)\mathrm{sin}(\theta)$ (dotted line) and a model with line nodes where  $\Delta_k=\Delta(T)\mathrm{cos}(\theta)$ (dashed-dotted line). Here $\Delta(T)$ is the gap temperature dependence from Ref.~\onlinecite{ProzorovMagPen} with $\Delta(0)=1.6k_BT_c$ and $3.5k_BT_c$ for the respective models. The presence of multiple electron and hole Fermi surface sheets revealed by band structure calculations \cite{138doi:10.1143/JPSJ.81.103704,139PhysRevB.86.174507}, as well as  a gap significantly smaller than the BCS value derived from fitting $\Delta\lambda(T)$ at low temperatures, suggest the possibility of multi-gap superconductivity. Therefore  the data are analyzed using a two-band $\gamma$ model \cite{57PhysRevB.80.014507} following Ref.~\onlinecite{57PhysRevB.80.014507}, where the superconducting gap on each band is calculated self-consistently using the Eilenberger quasi-classic formulation. The parameters are the partial density of states $n_1$ and $n_2$, the  intraband pairing potentials $\lambda_{11}$ and $\lambda_{22}$ along with $\lambda_{12}$ and $\lambda_{21}$ which characterize the  interband coupling. The superfluid density of a two-band superconductor can be summarized as $\rho_s(T) = x\rho_1(T) + (1-x)\rho_2(T)$, where $\rho_i(T)$ is the single band superfluid density for the gap $\Delta_i(T)$ ($i=1,2$) and $x$ is the relative weight of $\rho_1(T)$ \cite{ProzorovMagPen}. When using this procedure to fit the data with $\lambda_{12}=\lambda_{21}$ , the free parameters are $n_{1}$, $\lambda_{11}$, $\lambda_{12}$, $\lambda_{22}$ and $x$. We obtain a good fit across the whole temperature range with the best fitting parameters of $n_{1}$=0.4, $\lambda_{11}$=0.25, $\lambda_{22}$=0.153, $\lambda_{12}$=0.016 and $x$=0.43. The value of $x$ is close to $n_1$, suggesting the Fermi velocities of each band are similar. The fit to the $\gamma$~model is shown by the solid line in Fig.~\ref{fig2} and the zero temperature gap magnitudes are $\Delta_1(0)=1.29k_BT_c$ and $\Delta_2(0)=2.04k_BT_c$. The smaller gap value agrees well with $\Delta(0)=1.30k_BT_c$ obtained from fitting $\Delta\lambda(T)$  (Fig.~\ref{fig1}), as expected for two-band superconductors \cite{twobandrev}.

\begin{figure}[t]
\centering
\includegraphics[width=8.0cm]{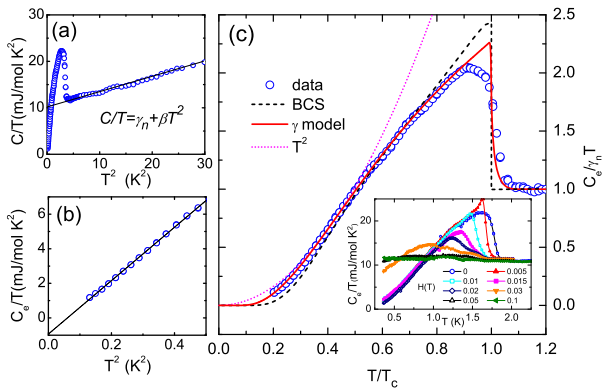}
\caption{(Color online) (a) Specific heat $C/T$ against $T^2$ for LaNiGa$_2$. The solid line shows a linear fit above $T_c$. (b) The electronic contribution to the specific heat $C_{e}/T$ against $T^2$ at low temperatures. The solid line shows a linear fit which extrapolates to negative $C_{e}/T$.  (c) Temperature dependence of $C_{e}/T$, normalized to the normal state value. The solid line shows the $\gamma$ model fit, while the dashed and dotted lines show the behavior of a weakly coupled, isotropic BCS superconductor and a $T^2$ dependence respectively. The inset shows $C_{e}/T(T)$ in various applied magnetic fields} \label{fig3}
\end{figure}

Specific heat ($C$) results for LaNiGa$_2$ are shown in Fig.~\ref{fig3}(a), where $C/T$  follows a $T^2$ dependence above $T_c$. The normal state behavior is fitted with $C/T=\gamma_n + \beta T^2$, giving an electronic specific heat coefficient  $\gamma_n=10.54~$mJ/mol~K$^2$ and a Debye temperature  $\Theta_D=294$~K from the phonon contribution $\beta T^2$, consistent with previous results \cite{136PhysRevB.66.092503}. The electronic contribution of the specific heat ($C_e$) is obtained by subtracting the phonon term, and the temperature dependence of $C_e/\gamma_nT_c$ is shown in  Fig.~\ref{fig3}(c).  The dashed line shows the specific heat of an isotropic, weakly coupled BCS superconductor, which also deviates from the data. Although $C_e(T)/T$ is not saturated down to 0.35~K, this is consistent with the penetration depth measurements, where low-energy quasiparticle excitations are present down to around 0.25~K. While it can be seen in Fig.~\ref{fig3}(b) that $C_e/T$ shows quadratic-like behavior at low temperatures, a negative value of $C_{e}(0)/T=-0.93$~mJ/mol K$^2$ is obtained upon extrapolating to zero temperature, suggesting a nodeless superconducting gap. $C_e(T)/T$ in the superconducting state is fitted using a two band model \cite{specHeattwoband},  $C_e(T)/T=xC_e^{\Delta_1}(T)/T+(1-x)C_e^{\Delta_2}(T)/T$, where $C_e^{\Delta_i}(T)/T$ is the single band electronic specific heat with $\Delta_i(T)$, calculated using the same expression as for the superfluid density fitting. From Fig.~\ref{fig3}(c), it can be seen that the data is well described by this model with fitted parameters  $n_{1}$=0.4, $\lambda_{11}$=0.261, $\lambda_{22}$=0.149, $\lambda_{12}$=0.02 and $x$=0.31. The derived specific heat jump is $\Delta C/\gamma T_c~=~1.28$ and the gap values at zero temperature are $\Delta_1(0)~=~1.08k_BT_c$ and $\Delta_2(0)~=~2.06k_BT_c$. This demonstrates that the specific heat measurements are consistent with two-gap superconductivity, as deduced from the superfluid density fitting.

\begin{figure}[t]
\centering
\includegraphics[width=7.0cm]{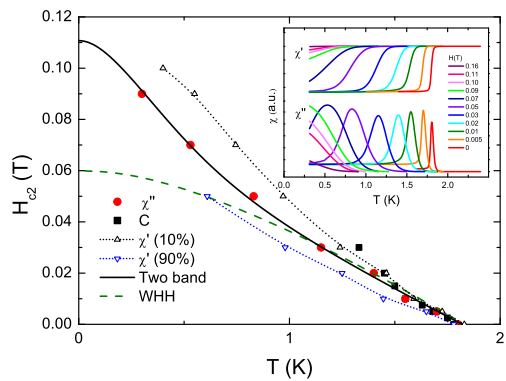}
\caption{(Color online) Upper critical field $H_{c2}(T)$ of LaNiGa$_2$ from specific heat and ac susceptibility measurements. For the ac susceptibility,  $H_{c2}(T)$ were obtained from the peak in $\chi''$, as well as where $\chi'$ reaches $10\%$ and $90\%$ of full screening, while the midpoint of the transition was used for the values from specific heat measurements.  The solid and dashed lines show the calculated values using a two-band and WHH model respectively, while the dotted lines are guides for the eye. The inset shows the temperature dependence of the real and imaginary parts of the ac susceptibility in various fields.} \label{fig4}
\end{figure}

To determine the upper critical field [$H_{c2}(T)$] of LaNiGa$_2$, we measured the ac susceptibility $\chi$ (inset of Fig.~\ref{fig4}) and specific heat (inset of Fig.~\ref{fig3}(c)) in various magnetic fields.  Note that a transition can not be clearly resolved in the specific heat data for applied fields greater than 0.03~T, the reason for which is not clear and requires further studies.  As shown in Fig.~\ref{fig4}, $H_{c2}(T)$ is almost linear near $T_c$. However, the curvature of $H_{c2}(T)$ shows a clear upturn at low temperatures, deviating from the Werthamer-Helfand-Hohenberg (WHH) model (dashed line) \cite{140PhysRev.147.295}. Such a negative curvature of $H_{c2}(T)$ is a common feature of multiband superconductivity. For a multiband system taking into account both interband and intraband couplings, $H_{c2}(T)$ can be calculated following Ref.~\onlinecite{141PhysRevB.67.184515}. The upper critical field was fitted with the same parameters used to fit the superfluid density, so that the only free parameters were the diffusivities of the  bands ($D_1$ and $D_2$). The data are well fitted by the model, as shown by the solid line in Fig.~\ref{fig4} and therefore $H_{c2}(T)$ is in good agreement with two-band superconductivity. The obtained value of $D_2/D_1$ is 0.15, while the extrapolated zero-temperature upper critical field $\mu_0H_{c2}(0)\simeq0.11~$~T. 

Therefore measurements of the penetration depth, specific heat and upper critical field consistently support two-gap superconductivity in LaNiGa$_2$, with both gaps being fully open. Similar two-gap behaviour was also observed in LaNiC$_2$\cite{NewJ.Phys.15.053005}, suggesting that nodeless, two-gap superconductivity is another common feature of these compounds, in addition to TRS breaking. In what follows we propose a unified view of these materials in which the two phenomena have a common origin.

Significant differences exist between the two materials. Electronic structure calculations reveal that either one or two bands cross the Fermi level ($E_{F}$) in LaNiC$_2$ \cite{JPSJ.78.084724,PhysRevB.80.092506}, while LaNiGa$_2$ has a very different Fermi surface, with several bands at $E_F$\cite{138doi:10.1143/JPSJ.81.103704,139PhysRevB.86.174507}. Moreover, whereas the crystal structure of $\mbox{LaNiGa}_{2}$ has a center of inversion, that of $\mbox{LaNiC}_{2}$ lacks it. As a result, in $\mbox{LaNiC}_{2}$ spin-orbit coupling lifts the spin degeneracy of the conduction bands and the superconducting state may be a mixture of spin-singlet and spin-triplet components \cite{NCSGorkov,FrigNCS,HQYNCS}. In contrast, for $\mbox{LaNiGa}_{2}$ such a state is forbidden by symmetry.

Several works have discussed $\mbox{LaNiC}_{2}$ and $\mbox{LaNiGa}_{2}$ in terms of a conventional BCS pairing mechanism \cite{PhysRevB.80.092506,138doi:10.1143/JPSJ.81.103704,139PhysRevB.86.174507,APL104022603} and this scenario  leads to fully-gapped superconductivity with two-gap behavior arising from the involvement of two distinct bands. However, such theories are not readily reconciled with the observation of TRS breaking in $\mu$SR measurements of both compounds \cite{135PhysRevLett.102.117007,137PhysRevLett.109.097001}.  To address this, it has been proposed that in $\mbox{LaNiC}_{2}$  the broken TRS may arise from a small admixture of triplet pairing to an otherwise largely conventional superconducting order parameter \cite{PhysRevB.80.092506}. Alternatively, it has been hypothesized that a non-trivial phase factor between the $s$-wave gaps in  two different bands (an $s+is$ state) might be responsible for broken TRS in either material \cite{138doi:10.1143/JPSJ.81.103704}.  However, the point groups of the crystal structures of  $\mbox{LaNiC}_{2}$ and  $\mbox{LaNiGa}_{2}$ are $C_{2v}$ and $D_{2h}$, respectively, both of which only have one-dimensional irreducible representations \cite{135PhysRevLett.102.117007,137PhysRevLett.109.097001} whereas a multi-dimensional order parameter is required to break TRS at $T_{c}$ \cite{TRSSig}. In the triplet admixture scenario, the relevant point group is the double group $C_{2v,J}$ whose irreducible representations have the same dimensionality as those of $C_{2v}$. In the $s+is$ scenario, the point group is either $C_{2v}$ or $D_{2h}$, if the bands are strongly-coupled, or the products $C_{2v}\otimes C_{2v}$ or $D_{2h}\otimes D_{2h}$, if they are decoupled, which also only have one-dimensional  irreducible representations. Thus in either scenario the broken TRS would require a first-order transition or multiple superconducting phase transitions \cite{PhysRevB.82.174511,PRL772284}. While the latter has been  observed in some superconductors such as UPt$_3$ \cite{UPt3PRL}, there is no experimental evidence in these compounds.

In the case of weak spin-orbit coupling the relevant point groups for $\mbox{LaNiC}_{2}$ and $\mbox{LaNiGa}_{2}$ are $C_{2v}\otimes SO(3)$ and $D_{2h}\otimes SO(3)$,
respectively, both of which have three-dimensional irreducible representations and there are four TRS-breaking superconducting instabilities \cite{135PhysRevLett.102.117007,137PhysRevLett.109.097001,PhysRevB.82.174511}, all of them in the purely-triplet channel and thus
in stark contrast to the above scenarios. All four instabilities correspond to nonunitary (equal-spin) pairing, for which we expect an additional, sub-dominant order parameter, in the form of a bulk magnetisation appearing below $T_{c}$, which may have been observed in $\mbox{LaNiC}_{2}$ \cite{LaNiC2Mag}. However, all of these nonunitary triplet states have nodal gap functions, which is clearly inconsistent with this work and Ref.~\onlinecite{NewJ.Phys.15.053005}.

As a result, we suggest that a new mechanism may be present in these materials. An isotropic gap which does not change sign can result from an on-site interaction, which is not possible for equal spin pairing in a single-orbital model, but could result from a local attraction between electrons with equal spins on different orbitals. The pairing potential has the form $\Delta_{\alpha,\beta}^{n,m}(\mathbf{k})$, where $n,m$ are orbital indices and $\alpha,\beta$ are the spin indices of the two paired electrons. For an isotropic gap with the formation of Cooper pairs within one orbital, $\Delta_{\alpha,\beta}^{n,m}(\mathbf{k})=\Delta_{\alpha,\beta}^{n,n}$ and therefore to keep the gap function antisymmetric under the exchange of two fermions, it is necessary that $\Delta_{\alpha,\beta}^{n,m}=-\Delta_{\beta,\alpha}^{n,m}$, that is there is singlet pairing between electrons of opposite spins. However if the pairing occurs between electrons on different orbitals, the condition for triplet pairing $\Delta_{\alpha,\beta}^{n,m}=\Delta_{\beta,\alpha}^{n,m}$ can be met if $\Delta_{\alpha,\beta}^{n,m}=-\Delta_{\alpha,\beta}^{m,n}$, that is the change of signs is achieved through an antisymmetric orbital index. Similar scenarios have been proposed to make $d$-wave pairing and fully-gapped behaviour compatible in the iron pnictides \cite{arxivOng} and to propose fully-gapped triplet pairing in that same family of materials  \cite{PRL101057008}. Our approach generalises the work of Ref.~\onlinecite{PRL101057008} to the nonunitary case, allowing for broken TRS.

The simplest theory embodying the above ideas features an electron-electron interaction  $\hat{V}=-U\sum_{j,\sigma}c_{Aj\sigma}^{\dagger}c_{Bj\sigma}^{\dagger}c_{Bj\sigma}c_{Aj\sigma}$. Here $c_{Aj\sigma}^{\dagger}$ creates an electron in an $A$ orbital on the $j{\textrm{th}}$ lattice site with spin index $\sigma$. $\hat{V}$ describes attraction, of strength $U$, between two electrons with parallel spins that occupy different orbitals $A,B$ on the same site $j.$ Within a standard variational mean field theory, the effect of  $\hat{V}$ can
be described by two mean fields $\Delta_{\uparrow\uparrow}c_{Aj\uparrow}^{\dagger}c_{Bj\uparrow}^{\dagger}+\Delta_{\downarrow\downarrow}c_{Aj\downarrow}^{\dagger}c_{Bj\downarrow}^{\dagger}+\mbox{H.c.}$
and $\Phi_{A\sigma}c_{Aj\sigma}^{\dagger}c_{Aj\sigma}+\Phi_{B\sigma}c_{Bj\sigma}^{\dagger}c_{Bj\sigma}.$
$\Delta_{\uparrow\uparrow}$ and $\Delta_{\downarrow\downarrow}$ describe uniform, equal-spin pairing between an electron in an $A$ orbital and an electron in a $B$ orbital on the same site. Our pairing potentials correspond to the $\Delta_{1}$ and $\Delta_{-1}$ terms in Ref.~\onlinecite{PRL101057008}, while we do not include the $\Delta_{0}$ term. The additional mean
field $\Phi_{n\sigma}$ takes care of the spontaneous spin polarisation \cite{137PhysRevLett.109.097001,Miyake2014}.

Diagonalising the Bogoliubov-de Gennes equations yields low-energy quasiparticles which have a well-defined spin index $\sigma$ but mixed orbital character. For each value of the spin, the quasiparticle spectrum has four branches and depends on the details of the splitting between the $A$ and $B$ orbital energy levels and band hybridisations. Neglecting these, it simplifies to two doubly-degenerate branches $E_{\sigma}=\pm\sqrt{\left(\epsilon-\mu+\Phi_{\sigma}\right)^{2}+\left|\Delta_{\sigma\sigma}\right|^{2}}$, which yields two fully open gaps of different sizes for $\uparrow\uparrow$ and $\downarrow\downarrow$ pairing. Such a simple model is consistent with electronic structure calculations of LaNiGa$_2$, which reveal the presence of two pairs of Fermi surface sheets, which are in close proximity in the Brillouin zone \cite{139PhysRevB.86.174507}. The details of the derivation and more general expressions will be provided elsewhere. Spectroscopically, this could be very similar to the conventional two-band behaviour captured by the $\gamma-$model used to fit our data. However, note that the two values of the gap are associated with two different values of the spin, rather than two band indices.

Further hints of an unconventional pairing mechanism come from recent measurements of LaNiC$_2$ under pressure, which reveal a broad superconducting dome, where the maximum of $T_c$ coincides with a crossover from a metallic normal state to one with strongly correlated electronic interactions \cite{PhysRevB.90.220508}. One possibility is that fluctuations of the correlated state mediate the pairing interaction, which might then look quite different from the simple, on-site form used above. Alternatively, the local attraction between equal spins could result from Hunds rules. Furthermore, our theory provides a mechanism for an on-site attraction leading to triplet pairing, suggesting the possibility of TRS breaking superconductivity mediated by phonons.

To summarize, we have performed measurements of London penetration depth, specific heat and upper critical field which indicate two-gap, nodeless superconductivity in LaNiGa$_2$. The presence of two gaps in both LaNiGa$_2$ and LaNiC$_2$ allows us to propose a novel non-unitary triplet state, where the gap symmetry has even parity. This can reconcile the observation of  fully gapped behavior and the breaking of TRS in both compounds and further work is required to elucidate the mechanism which leads to this novel pairing state.

\begin{acknowledgments}
This work was supported by the National Natural Science Foundation of China (No.11474251), the Fundamental Research Funds for the Central Universities, the National Basic Research Program of China (No.2011CBA00103) and the Max Planck Society under the auspices of the Max Planck Partner Group of the MPI for Chemical Physics of Solids, Dresden.
\end{acknowledgments}

\end{document}